\documentclass[12pt,preprint]{aastex}
\usepackage{amsmath, amsthm}

\begin{document}

\title{Transrelativistic Synchrotron Emissivity, \\
Cross-Section, and Polarization}
\author{Brandon Wolfe$^{1}$ and Fulvio Melia$^{1,2}$}
\affil{$^1$ Physics Department, The University of Arizona, Tucson, AZ 85721}
\affil{$^2$ Steward Observatory, The University of Arizona, Tucson, AZ 85721}

\begin{abstract}
The spectrum and polarization produced by particles spiraling in a magnetic
field undergo dramatic changes as the emitters transition from nonrelativistic
to relativistic energies.  However, none of the currently available methods
for calculating the characteristics of this radiation field are adequate for
the purpose of sustaining accuracy and speed of computation in the intensity,
and none even attempt to provide a means of determining the polarization fraction 
other than in the cyclotron or synchrotron limits.  But the transrelativistic
regime, which we here find to lie between $5\times 10^7$ K and $5\times 10^9$ K
for a thermal plasma, is becoming increasingly important in high-energy
astrophysical environments, such as in the intra-cluster medium, and in the
accretion flows of supermassive black holes. In this paper, we present simple,
yet highly accurate, fitting formulae for the magnetobremsstrahlung (also
known as cyclo-synchrotron) emissivity, its polarization fraction (and content), 
and the absorption cross-section. We demonstrate that both the harmonic and
high-energy limiting behavior are well represented, incurring at most an
error of $\sim 5\%$ throughout the transition region.  
\end{abstract}

\keywords{acceleration of particles --- magnetic fields --- plasmas --- radiation 
mechanisms: non-thermal --- radiation mechanisms: thermal --- relativity}

\section{Introduction}

An electron circling a constant magnetic field at $50 \%$ the speed of light emits about 
$15$ times less energy in a given time than is predicted by the synchrotron theory. It is 
similarly less likely to re-absorb radiation, and will emit light with a polarization 
fraction three-halves the amount predicted by that theory. Yet, while the foundation for 
a complete theory of transrelativistic magnetobremsstrahlung (or cyclo-synchrotron radiation) 
was put forward some time ago (Oster 1963), the solution of this equation has been given only 
recently, and then only partially (see, e.g., Petrosian 1981; Takahara and Tsuruta 1982; Melia 
1994; Mahadevan et al. 1996). Here, we provide analytic fits to the (complete) numerical solution, 
and demonstrate the power of this technique for thermal and pure power-law (mildly relativistic) 
particle distributions. The application to arbitrary particle populations is equally straightforward.

Quasi-relativistic and relativistic particles are quite common in high-energy astrophysical
environments. Plasmas that contain energetic electrons also tend to be magnetized. Often,
the dominant emission mechanism in these sources is thermal or nonthermal cyclo-synchrotron
radiation.  For example, this type of situation is increasingly seen in hot accretion
flows onto compact objects, such as weakly accreting supermassive black holes (see also
Melia 1994; Melia and Falcke 2001). The frequency and emissivity of the radiation undergo dramatic
changes as the particle energy transitions from nonrelativistic to extreme relativistic
values. The ever greater smearing of successive harmonics causes a pronounced
overlapping that forms an almost monotonically decreasing spectrum at high temperature,
whereas the lower-energy spectrum consists of discrete lines of rapidly decreasing intensity
separated by the cyclotron (or magnetic) frequency $\nu_b\equiv\omega_b/2\pi= 2\pi\, e B/m_e c$, 
in terms of the magnetic field $B$.

A problem of this significance has attracted a great deal of attention.  Its provenance dates
to Schwinger (1949), who calculated the emission in the extreme nonrelativistic limit. In this
case, the frequency of emitted light is precisely the inverse of the amount of time it takes
an electron to make one orbit, and integer multiples of this fundamental magnetic frequency,
$\nu_b$.  In Oster's (1961) formulation, the field equations of a particle circling a constant
magnetic field are derived from the Lienard-Wiechert potentials and Fourier-analyzed at the
retarded time (see, e.g., Melia 2001).  The cyclotron harmonics are therefore broadened, 
not due to Doppler shifts but to the relativistic dependence of the resonance frequency on 
the particle energy, until they reach a synchrotron continuum. 

The language of this field reflects its long history. `Cyclotron' is nonrelativistic emission
($T<10^7$ K) which is symmetric, given by keeping only the first term in the series expansion 
of the Bessel function in Equation (2) (below). `Gyroresonance' or `gyrosynchrotron' is the 
slightly relativistic ($T \sim 5\times 10^7$ K) emission characterized as an asymmetric dipole, 
and given by taking a Carlini expansion of the Bessel function. `Synchrotron' is fully relativistic 
($T > 5\times 10^9$ K), continuous, and radially asymmetric; it is calculated from the general 
formula by approximating the Bessel functions using Airy functions. The entire range of emission 
is sometimes called `Magnetobremsstrahlung', implying the scattering of virtual photons in the
magnetic field by the gyrating charges. 

An analytic approach to Oster's complete description was made by Petrosian (1981) in the 
mildly relativistic regime $\nu / \nu_b \gg 1$, $\gamma > 1$.  This motivated a series of 
calculations of the emissivity, cross-section, and polarization, summarized in Dulk (1985),
who simplified the expressions and applied them to solar radio emission.  Ghisellini 
\& Svensson (1991) provided angle-averaged expressions of the Petrosian approximation for 
emission and absorption.  Melia (1994) then presented a numerical solution from Oster's
formulation, which was limited to an observer angle of $\pi/2$. The first complete numerical 
solution was put forward by Mahadevan, Narayan, and Yi (1996, MNY96), and this was followed 
by a detailed fitting of the numerical solution by Marcowith and Malzac (2003, MM03).

Each of these approaches leaves significant room for improvement (see, e.g., Fig. 1). 
The limitation of observer angle in Melia (1994) neglects a significant portion of the emissivity. 
MM03 consider only the first $10$ harmonics, a procedure that fails entirely in the region 
above $\nu/ \nu_b > 10$---a surprisingly high number of harmonics are required to correctly 
transition to the relativistic regime (see Fig. 2).  And the approximations of Petrosian (1981) 
and Ghisellini \& Svensson (1991) do not capture the harmonic quality of transrelativistic 
emission; the former fares less well in the low-frequency limit, the latter in the low-velocity 
limit. The correction needed around $\beta\equiv v/c = 3/4$ is $\sim 3$, while around 
$\beta = 1/4$ it is $\sim 7$.  The Carlini expansion is an excellent approximation, but it 
nevertheless fails below $\nu \sim 6\, \nu_b$ (Petrosian and McTiernan 1984). 
The calculation of MNY96 is complete, but their fits only apply 
to four specific thermal luminosities.  Their result is irretrievably numerical, and cannot be 
reproduced or generalized without duplicating the full (time intensive) integration.

Despite these shortcomings, there is no need to attempt an additional solution if it is not 
simpler to use, as well as more accurate. The polynomial fits we present here require a small 
table of coefficients, yet capture both the scale and harmonic nature of radiation in the 
transrelativistic regime. In the next section, we describe the calculational procedure, and
then apply it to a determination of the emissivity and polarization fraction of the radiation
in \S\S\ 3 and 4, respectively. It is also straightforward to calculate the synchrotron absorptive
cross section, which we do in \S\ 5, and we present our conclusions in \S\ 6.

\section{Calculation}
The complete emissivity (energy per unit volume, per unit time, per unit frequency)
of a particle distribution $n(\gamma)$ with (dimensionless) velocity components 
$\beta_\parallel(\gamma)$ and $\beta_\perp(\gamma)$ relative to the underlying magnetic 
field, is given by 
\begin{equation}
\frac{dE}{dV\,dt\,d\nu} = \int^\infty_1 d\gamma\; n(\gamma)\;j_\nu(\gamma)\;,
\end{equation}
in terms of the single particle emissivity
\begin{align}
j_\nu &\equiv \bigg( \frac{e^2 \omega_b}{2 \pi c} \bigg)
               \int^{\pi/2}_0 d\theta_p\;\sin\theta_p 
               \int^\pi_0 d\theta\;\sin\theta\, 
               \frac{f(\chi)\,\chi^2}{1 - \beta_\parallel \cos\theta}\;\times\nonumber \\
               &\bigg[ \sum^{990}_{m=1} 
               \bigg( \frac{\cos\theta - \beta_\parallel}{\sin\theta} \bigg)^2
               J_m^2 \bigg( \gamma \chi \beta_\perp \sin\theta \bigg)
               + \beta_\perp^2 J_m^{\prime 2}
               \bigg( \gamma \chi \beta_\perp\sin\theta \bigg) \bigg]\;.
\end{align} 
In this expression, $\chi = \nu/\nu_b$ is the ratio of the observed frequency to the first 
harmonic, $\theta_p$ is the angle between the velocity and the line-of-sight, and $\theta$ 
is the angle between the particle's velocity and the magnetic field.  The function
\begin{equation}
f(\chi) = \frac{15}{16 \Delta \chi} \bigg[ 1 - \bigg( \frac{2}{\Delta \chi^2} \bigg)
        (\chi - \chi_c)^2 + \bigg( \frac{1}{\Delta \chi^4} \bigg) (\chi - \chi_c)^4 \bigg]
\end{equation} 
is a computational approximation to the delta function at the harmonic resonances (see MNY96).

Note here that the sum extends to the 990th harmonic, imposed by computational precision,
and that a typo of the Bessel function's argument from MNY96 has been corrected. The term 
proportional to the square of the Bessel function $J_m$ is emitted with ordinary (O) polarization, 
while the term proportional to the square of its derivative is associated with extraordinary (X) 
polarization.

The comparison between previous approximations to the solution of $j_\nu$ can best be made with
reference to Figure~1, which shows $j_\nu$ as a function of $\nu/\nu_b$ for $\gamma=1.5$.
Here and throughout we plot the dimensionless emissivity $\phi_\nu = j_\nu (c/e^2 \nu_b)$.  As discussed
in the introduction, and shown more graphically here, none of the previous approaches can simultaneously
handle both the harmonic structure at low energy and the behavior at high energy. We also see immediately
(Fig.~2) that a very large number (here 990) of harmonics must be used to adequately represent $j_\nu$ for
$\nu/\nu_b> 10$. 

To map the transition of the emissivity from nonrelativistic to highly relativistic particle dynamics, we 
show in Figure~3 other values of the integral $j_\nu$ for a series of velocities $\beta$, together with the 
numerical evaluation of Equation (2) and our fitting function, described below. On this scale, the
complete solutions and their fits are indistinguishable. However, neither the cyclotron nor the
synchrotron limits are adequate representations for these mildly relativistic values of $\beta$.

To make these fits, we restrict our attention to $-1.0 < \log \chi < 2.0$, where most of the 
transrelativistic emission takes place. Our ansatz takes the form
\begin{equation}
j( \nu, \gamma) = \sum_m^{Fejer} \bigg[ \sum_n C_{m,n}\, \log_{10} \gamma \bigg] J_1 \left(\frac{\alpha_m}{L} 
\,\log_{10} \chi\right)
\end{equation}
(see Appendix A for a definition of the Fejer-averaged sum). The terms within brackets are the constants of 
the initial emissivity expansion, which have themselves been re-expanded as functions of $\gamma$. The constants 
$C_{m,n}$ then form a $60\times 25$ table of numbers that completely specify the emissivity. 

The fit is valid between $0.1 < \beta < 0.98$, beyond which the cyclotron and synchrotron formulae apply. 
It has a mean error of between $3$--$7 \%$ in order (shown in Figure~4), the largest error lying between
harmonics near the cyclotron limit. By comparison, the approximation of Petrosian (1981) fails in these
locations by several orders.  The table $C_{m,n}$, together with a function that returns the emissivity for any 
given $\beta$, is available in the electronic edition of this paper.

\section{Particle Averaged Emissivity}
The evaluation of Equation (1) for the particle averaged emissivity is straightforward
for any distribution $n(\gamma)$. For illustrative purposes, we here consider the
special case of an arbitrarily relativistic thermal (equilibrium) distribution,
\begin{equation}
n(\gamma)\,d\gamma = \frac{\psi}{K_2(1/\psi)}\,\beta \gamma^2 \exp\bigg(-\frac{\gamma}{\psi} \bigg)\, d\gamma\;,
\end{equation}
where $\psi \equiv kT/m_e c^2$, and that of a pure power law,
\begin{equation}
n(\gamma)\,d\gamma=n_0\,\gamma^{-p}\,d\gamma\;,
\end{equation}
arising, e.g., in shock acceleration, or stochastic particle acceleration (a second-order Fermi process)
in a turbulent magnetic field. 

The thermally averaged synchrotron emissivity from our
expansion is shown in Figure~5, compared to the full numerical result. When $\beta$ extends
past $0.97$ we use Petrosian's approximation. Both the low- and high-frequency
behaviors are well matched, as is the harmonic structure. The mean error incurred with the use of
our expansion is never bigger than $\sim 5\%$ over the entire temperature range $4\times 10^7$ K 
$<T<4\times 10^9$ K. Caution must be used with the angle-averaged synchrotron fit of MNY96, which
is not a good approximation in this regime (Figure~6).

It should be emphasized that the transition from relativistic to nonrelativistic synchrotron
emissivity occurs at a different temperature than that of the corresponding transition in the
particle distribution. This is due to the fact that the cyclotron limit applies when $m\beta \ll 1$, 
where $m$ is the harmonic number (Schwinger 1949), and not $\beta \ll 1$, as is usually thought. 
To make this point clearer, let us compare the thermally averaged energy per particle $E = \xi 
(3/2) kT$ (where $\xi\rightarrow 1$ in the nonrelativistic limit, and $\xi\rightarrow 2$ for fully
relativistic particles) to the thermally averaged magnetobremsstrahlung power.  The factor $\xi$
therefore defines the non-relativistic and ultra-relativisitc domains. It is $\xi>1$ 
above $T \approx 5\times 10^8$ K, and $\xi<2$ below
$T \approx 5\times 10^{10}$ K. The magnetobremsstahlung power, 
however, is dominated by the first cyclotron harmonic until $T \approx 1\times 10^8$ K, 
and is within a few percent of synchrotron by $T \approx 5\times 10^9$ K. (Equation 1 varies from 
the first cyclotron harmonic by a factor $\sim 2$, due to angle-averaging.)

That is, relativity's effect on cyclotron emission is
important at approximately one-tenth the temperature at which it becomes important for the distribution 
itself.  Our intuition of what constitutes a relativistic temperature must be modified in this case.

We also average over the power-law distribution. We plot the power-law averaged synchrotron 
emissivity in Figure~7.  
In all of these cases, the high-energy spectrum matches the synchrotron power relation, since power-laws
always extend to high $\gamma$. However, power-laws are also dominated by $\gamma \sim 1$, thus there
is a spectral bump near $\chi \sim 0.6$ for all $p<1.5$, and emission below the peak remains dominated 
by the broadened first harmonic, rather than an overlap of many harmonics as in the synchrotron regime. Therefore,
this region underproduces emission with respect to the synchrotron approximation; neither the synchrotron 
nor the Petrosian approximations apply below $\nu = \nu_b$ for any spectral index above $p=2$. 
In this region, we may relate the
spectral index $\alpha$ and the particle distribution index $p$ as
\begin{equation}
p=5.84\alpha - 5.19
\end{equation}
which is at odds with the synchrotron relation.

The expanded emissivity may be averaged over any particle distribution---power law, bi-thermal, 
or hybrid, as required---with equal facility.
   
\section{Polarization Fraction}
Non-relativistic cyclo-synchrotron polarization is straightforward: along the axis of rotation, light 
is circularly polarized in the direction of rotation, while in the plane of rotation it is linearly 
polarized. As light is beamed into a boosted direction, however, these orientations become mixed; thus 
the two limiting forms have quite different results.

The polarization fraction at $\theta = \pi/2$ was first solved by Trubnikov (1962).
Petrosian's approximation was followed by Robinson and Melrose (1984) and Dulk (1985)
for non-isotropic pitch angle distributions. Petrosian and McTiernan (1983) calculated 
the polarization fraction at the `critical energy' where synchrotron emission is peaked. 
In what follows, we refer to $\bar{\eta}_O/\bar{\eta}_X$ as the `polarization fraction',
and to $r = (\bar{\eta}_O-\bar{\eta}_X)/(\bar{\eta}_O+\bar{\eta}_X)$ as the `polarization content';
the distinction is required to compare with past results.

We have expanded the angle-integrated Equation (2) in both the ordinary and extraordinary polarization 
directions in the same manner as above, yielding two additional $65\times 20$ data tables. 
Figure~8 shows these numerical emissivities together with their expansion, and Figure~9 
shows the frequency-integrated single-particle polarization fraction $\bar{\eta}_O/\bar{\eta}_X$ 
from the nonrelativistic to the relativistic regime. In the low-energy limit, this fraction is 
dominated by low frequencies and moves correctly to the limit of $1/3$ given by Trubnikov, 
while in the high-energy limit it matches the synchrotron value. In between, the polarization 
fraction may be $3/2$ bigger than the synchrotron value. 

Figure~10 shows the polarization content, $r$, for a thermal distribution as a function of 
temperature and frequency.  The approximate expression for the thermal polarization fraction 
given by Petrosian and McTiernan (1983), valid at the critical energy and for $\psi\chi \gg 1$ is
\begin{equation}
 r = \cos(\theta) \bigg( \chi \psi\sin^2(\theta/6) \bigg)^{-1/3},
\end{equation}
and is plotted at $\theta = 79$ degrees (dashed).  Petrosian's result is therefore accurate near 
this angle, although it is not obvious \emph{a priori} what angle dominates emission in the 
transrelativistic regime (transrelativistic emission is not a simple dipole, see MNY96).
Both of these results are irreconcilable with those given by Dulk, however, perhaps because
his approximation includes an anisotropic pitch-angle distribution.

Our own approximation to the thermal polarization content, assuming isotropic pitch-angles, is 
\begin{equation}
r = \exp \bigg[ \log_{10}\chi \,(3.9\,\psi^{-1/3} -7.1\psi^{-1/4}+2.7) \bigg] (-1.6\,\psi+0.47), 
\end{equation}
which is valid between $-1<\log_{10}(\chi)<2$ and $0.01<\psi<2$.  It is accurate
to $40\%$ in the nonrelativistic case, where one should anyway use the cyclotron polarization, 
and $3\%$ once $\psi>0.1$.  A brief discussion of how this figure relates to the circular and 
linear polarization content is given in Appendix B.

\section{Cross Section}
Our analytic fits make the calculation of the synchrotron cross section particularly simple. 
We require
\begin{equation}
 \sigma_{Synch} (\nu, \gamma) = \frac{1}{2 m_e \nu^2} \frac{1}{\gamma^2 \beta} \frac{\partial}{\partial \gamma}
	\bigg[\gamma^2 \beta j(\nu \gamma) \bigg],
\end{equation}
where the term in brackets may be differentiated using the emissivity tables.  This cross-section 
is presented, along with the limiting forms of Ghisellini \& Svensson's (1991) Equation (23), and 
the synchrotron limit, in Figure~11.  At $\beta = 0.6$ in the low-frequency regime Petrosian's 
limit would incur an error of $\sim 2$ and the synchrotron limit an error of $\sim 1/3.5$.  By 
$\beta = 0.8$ these figures are $\sim 2.5$ and $\sim 1/2.5$, and the synchrotron limit may be 
favorably used above this. In the high-frequency regime, Petrosian's limit is always preferred.

\section{Conclusion}

In this paper, we have presented fitting formulae, appropriate for both analytic and high-speed numerical
calculations, for magnetobremsstrahlung in the transrelativistic regime. We have found that the
magnetobremsstrahlung transrelativistic regime actually lies between $T\sim5\times 10^7$ K 
and $T\sim 5\times 10^9$ K---lower in temperature than is usually thought on the basis of where
the transition occurs for the particle distribution itself. We have also concluded that for the latter 
half of this range, a harmonic number larger than $100$ is required for reasonable accuracy. We have
demonstrated the presence of a spectral bump for power-law distributions when $p>1.5$, and have
found a relation between spectral and distribution indices in a regime where neither the Petrosian 
nor the synchrotron approximations apply.
And, finally, we have shown that the polarization fraction of transrelativistic cyclo-synchrotron radiation is 
approximately $3/2$ that of its relativistic value.  

Thus, there is sufficient reason for those interested in standard thermal or power-law distributions 
in the transrelativistic regime to use the correct theory. High-energy plasmas are less likely to fall 
into these categories. 

Together with our previous work on a covariant kinetic theory (Wolfe and Melia 2005), these 
calculations form a coherent picture of high-energy magnetized plasmas. In upcoming papers,
we will apply these theories to several systems of current and active interest, including a
careful and accurate prediction of the $\gamma$-ray and neutrino production in the hierarchical 
merging of galaxy clusters, related to the non-thermal hard X-ray excess now detected from
these sources, and to weakly accreting supermassive black holes, such as Sgr A* at the
Galactic center, where plasma temperatures as high as $10^{11}$ K are reached within
$5-10$ Schwarzschild radii of the event horizon.  We anticipate that use of the polynomial
expansions we have presented here will extend to many other sub-branches of astrophysics and
to other physics disciplines where transrelativistic particle distributions are active.

\section{Appendix A: Fejer Averaging}

The Fourier-Bessel series expands the function $f(x)$---here the single-particle emissivity---as
\begin{equation}
f_m(x) = \sum_{r=1}^m s_r J_n(x \alpha_r/L)\;,
\end{equation}
where $\alpha_r$ is the $r$th positive root of $J_n$. Here $n$ is the order of the Bessel function, 
the coefficient $r$ is the expansion harmonic number
and $m$ is the largest expansion coefficient. The functions $f_m(x)$ and $f(x)$ are different for any $m<\infty$.

The Fejer averaging is a technique of signal processing which makes some non-convergent expansions converge,
and helps many expansions converge faster. Yet it is easy to apply.  This technique averages the 
contributions for increasing numbers of harmonics, 
\begin{equation}
F_m(x) = \frac{1}{m+1} \sum_{k=1}^m f_k(x),
\end{equation}
where the function $F_m(x)$ is now considered a more accurate version of $f(x)$ than $f_m(x)$.
However, we have found that extending the sum only to the higher half of the total expansion number
results in lower standard deviations. We denote Fejer sums as $\sum^{Fejer}$.

\section {Appendix B: Polarization Ellipse}

The emissivity in Equation (2) is expressed in ordinary and extraordinary modes. 
To place these in the observationally relevant linear and circular polarizations, we 
use a Poincare sphere with latitude T (the ratio of the ellipse's
major and semimajor axes) and longitude K (the ellipse's tilt angle). 
As we are interested only in the ratios of circular and linear polarizations,
we neglect the tilt angle K; this is equivalent to setting the Stokes
parameter $U$ to zero. Then the linear and circular polarization contents are
\begin{alignat}{2}
r_c = \frac{2 T_o}{T_o^2+1}r &\qquad r_l = \frac{T_o^2-1}{T_o^2+1}r\;,
\end{alignat}
where the ordinary and extraordinary axial ratios are
\begin{equation}
T_o = -\cos\theta ( a+\sqrt{1+a^2} )/ |\cos \theta| = -T_x^{-1}\;,
\end{equation}
and
\begin{equation}
a = \frac{ \sin^2 \theta }{ 2 \chi |\cos \theta|}\;.
\end{equation}
Thus the modes are circularly polarized ($T_o \approx -1, T_x \approx 1$) for $|\theta - 1/2\pi| \gg 1/2 \chi$
and linearly polarized ($T_o \approx -\infty, T_x \approx 0)$ for  $|\theta - 1/2\pi| \ll 1/2 \chi$.

\begin{figure}
\centering
\plotone{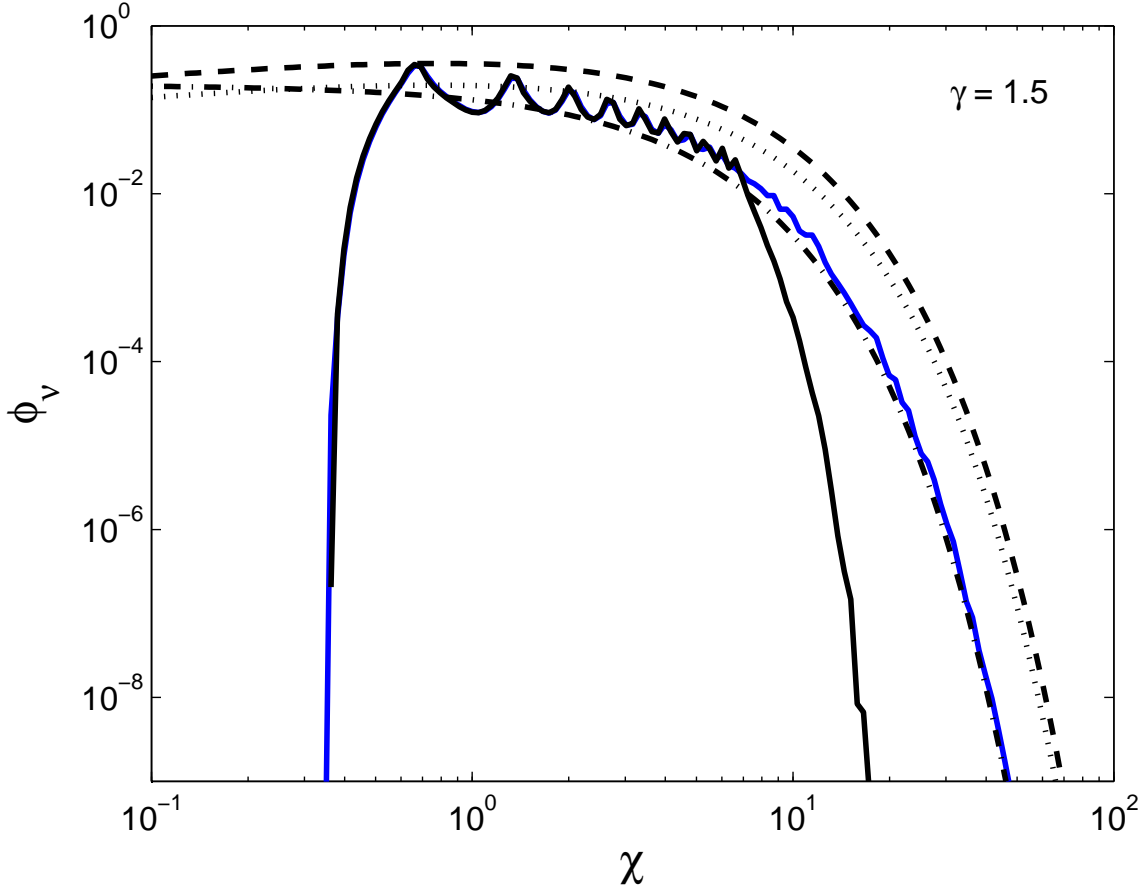}
\caption{
Several approximations exist for this historical problem. The dashed line is the angle-averaged 
synchrotron formula of Schwinger (1949). The dotted line is the analytic calculation in the 
relativistic regime of Ghisellini and Svensson (1991). The dash-dotted line is the approximation of 
Petrosian (1981). The two solid lines are our calculation, at $10$ harmonics (as used by MM03) 
and $990$ harmonics (which we will use throughout the paper).  Here and throughout we plot the 
dimensionless emissivity $\phi_\nu = j_\nu (c/e^2 \nu_b)$.
}
\end{figure}

\begin{figure}
\centering
\plotone{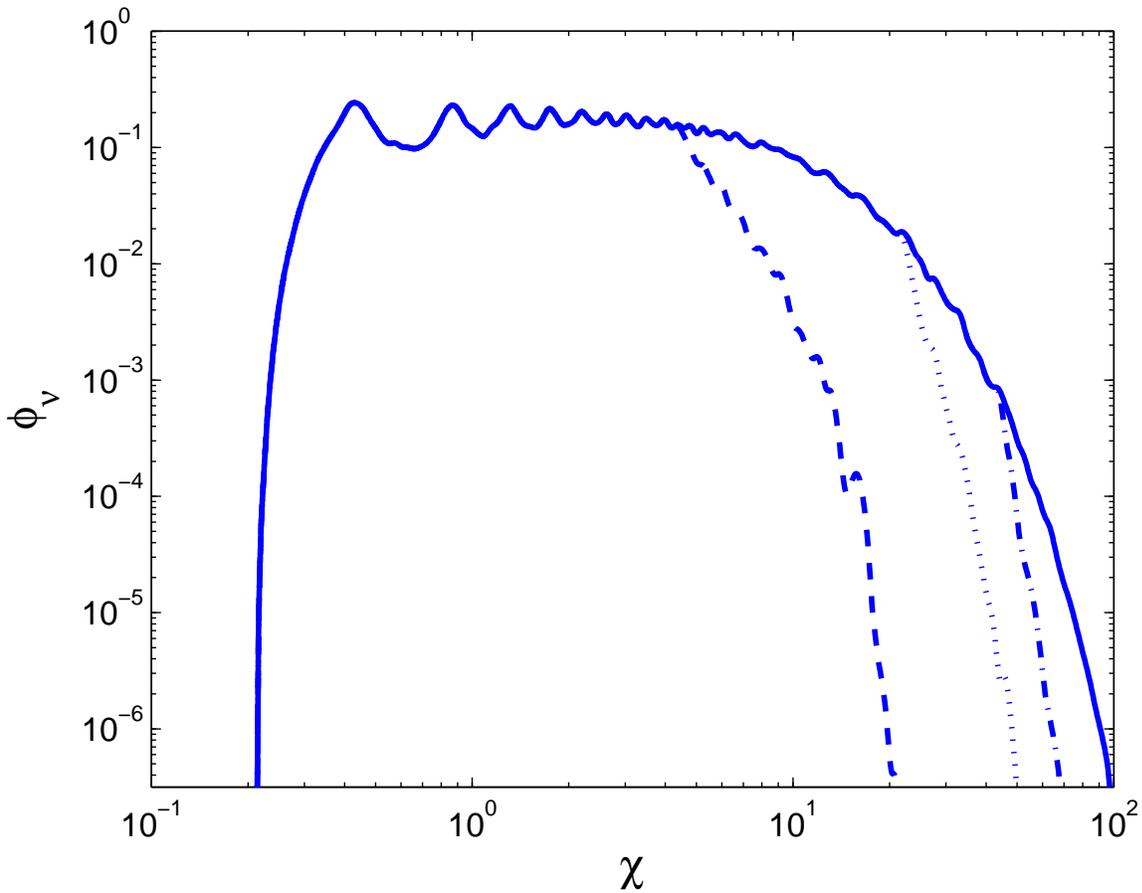}
\caption{
The emissivity of a particle with $\gamma=2$ changes as the harmonic 
number grows from $n=10$ (dashed), 50 (dotted), 100 (dash-dotted), to 300 (solid). The fundamental limitation
of machine precision on the highest allowed harmonic, which we find to be 990, makes integration to arbitrary
$\gamma$ impossible. However, any emission past $\log (\nu / \nu_b) = 2$ has already gone over to the synchrotron
limit; thus we limit our discussion to frequencies below this value throughout the paper. 
}
\end{figure}

\begin{figure}
\centering
\plotone{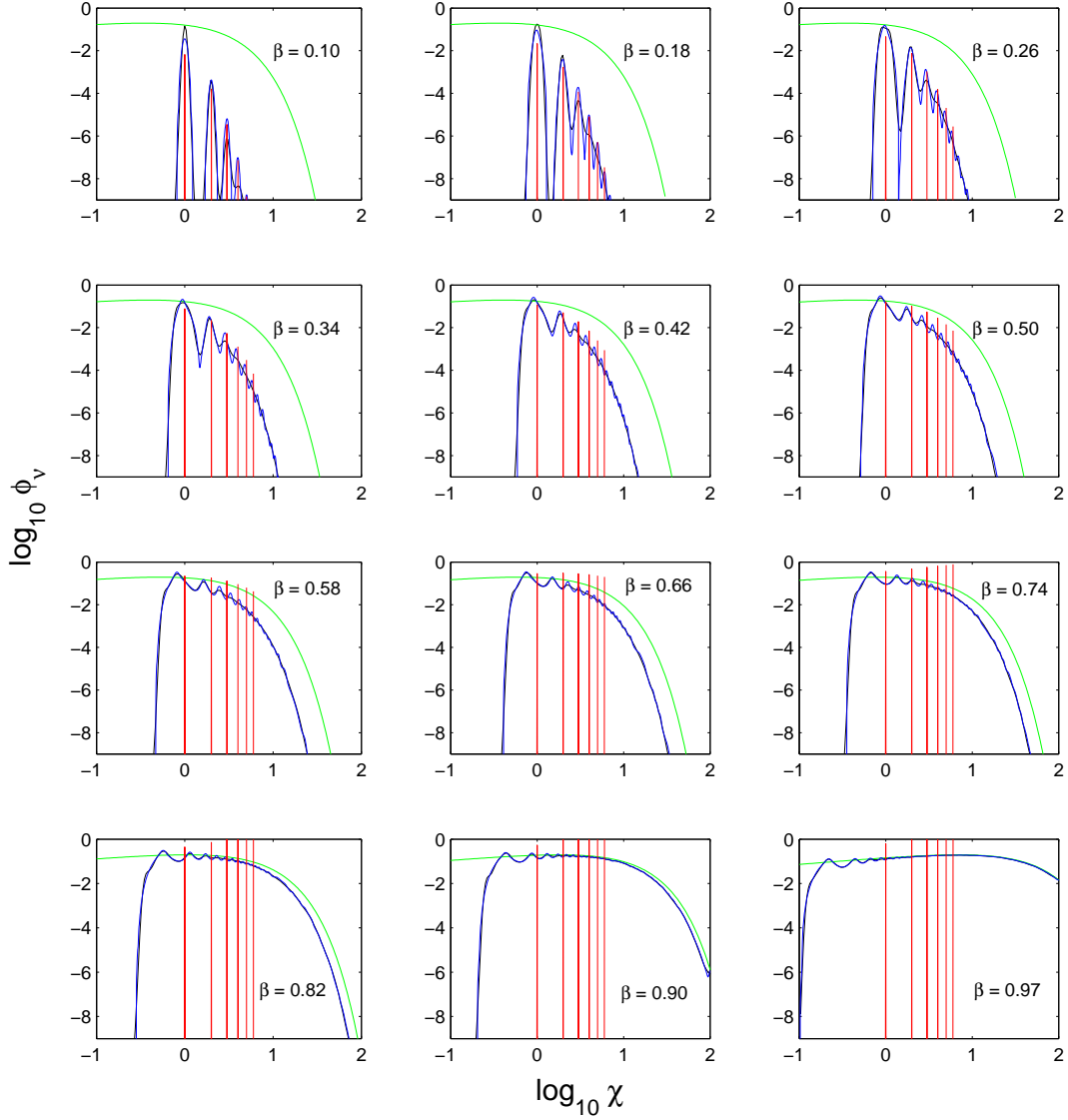}
\caption{
Here, cyclotron lines are broadened into a synchrotron continuum. The continuous line is the 
synchrotron limit of emissivity, the sharp harmonics the cyclotron limit. The numerical calculation
and its polynomial fit are generally indistinguishable; however note that in the
extreme nonrelativistic limit (top left), only two harmonics are sharply distinguished. 
The synchrotron limit only applies above $\gamma \sim 4$ (corresponding to $\beta=0.97$).
Here and throughout we plot the dimensionless emissivity $\phi_\nu = j_\nu (c/e^2 \nu_b)$.
}
\end{figure}

\begin{figure}
\centering
\plotone{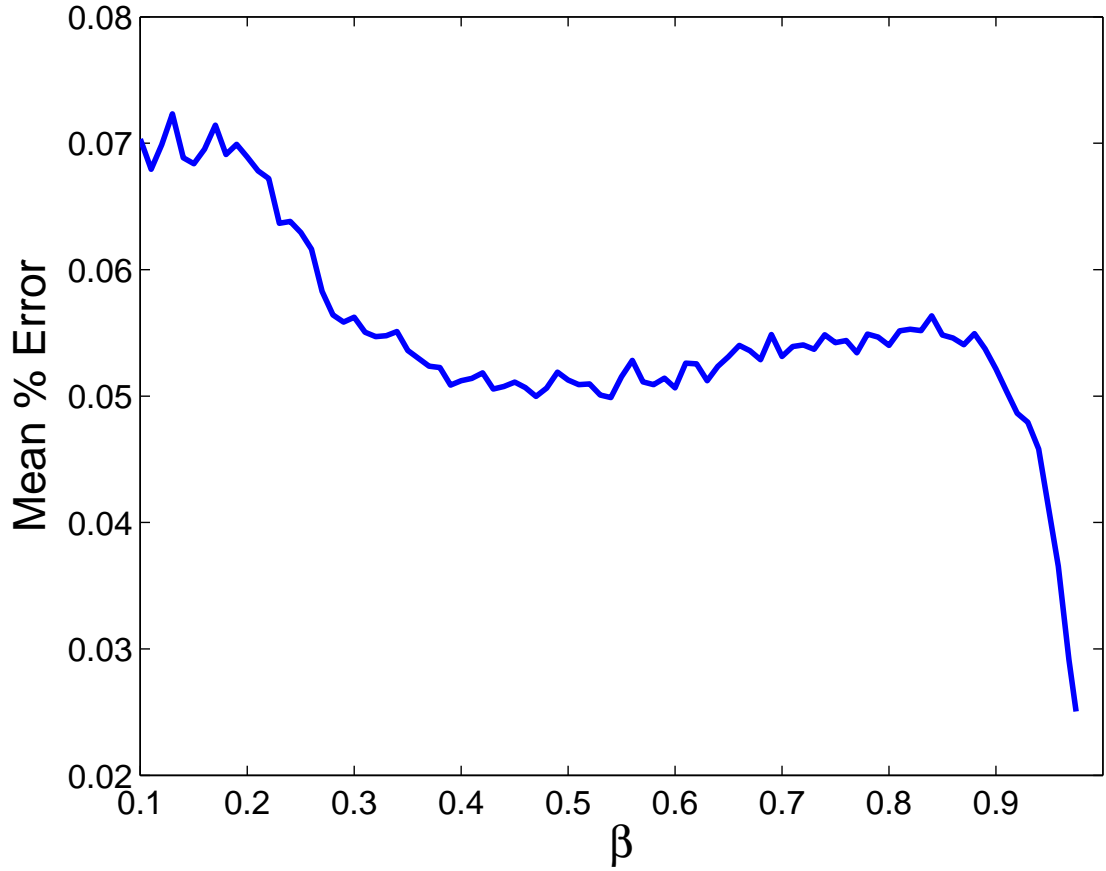}
\caption{
The mean error of our fit lies below $7 \%$ over the entire range $0.1 < \beta < 0.98$.
Error is worst ($\sim 10 \%$) between harmonics in the extreme nonrelativistic limit.
}
\end{figure}

\begin{figure}
\centering
\plotone{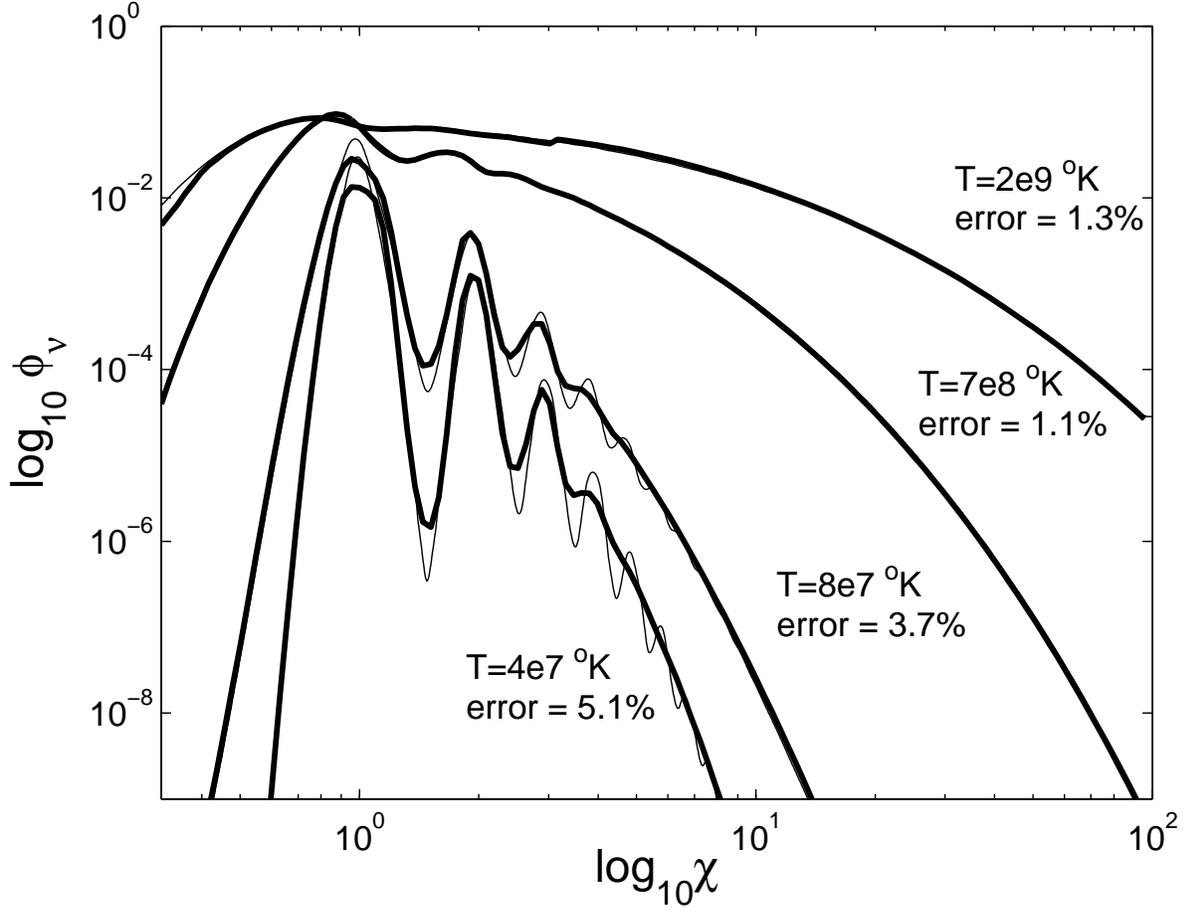}
\caption{
The thermally averaged synchrotron emissivities from our expansion (thick), as compared 
to the full numerical result (thin). Moving progressively upwards are $T=4\times 10^7$, 
$8\times 10^7$, $7\times 10^8$, and $2\times 10^9$ K.  The mean error of our fit is indicated.
}
\end{figure}

\begin{figure}
\centering
\plotone{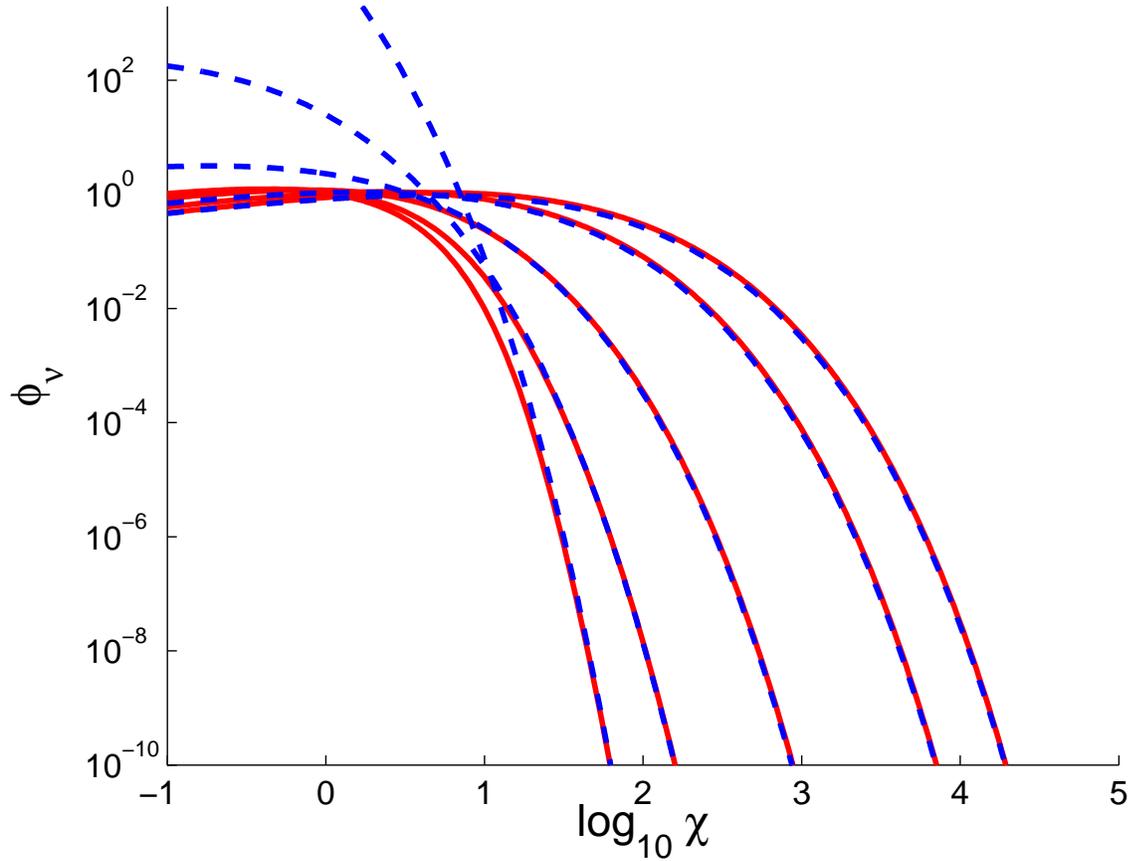}
\caption{MNY96 provide a single analytic fit (dashed) for the angle-averaged cyclo-synchrotron 
emissivity (Eq. 32), which fails in the synchrotron limit (solid) at mildly relativistic 
temperatures and low frequencies.  Of course, in this region it would be best to abandon the 
synchrotron functions entirely.
}
\end{figure}

\begin{figure}
\centering
\plotone{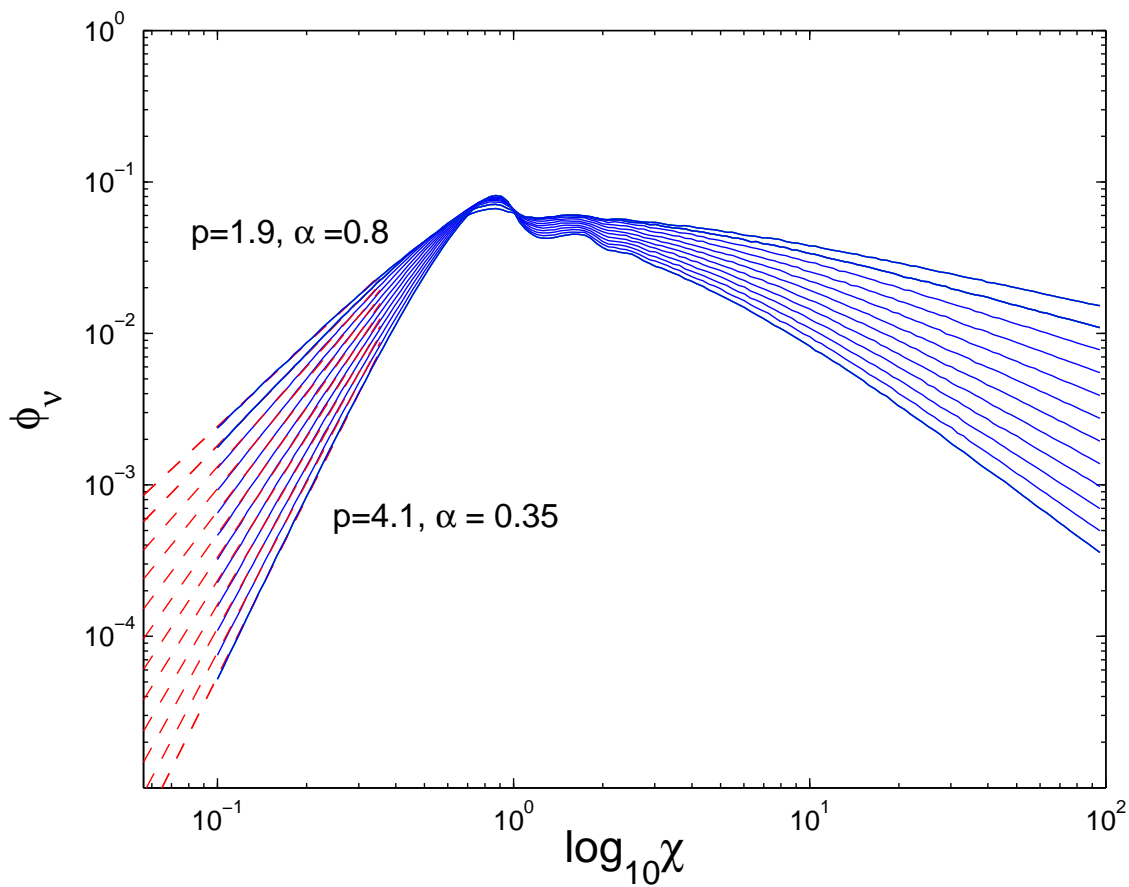}
\caption{
Cyclo-synchrotron emission from a power-law distribution of electrons, with $p$ moving progressively up
from $1.9$ (topmost) to $4.1$ (lowest) at spacings of $0.2$. The relation between distribution and spectral indices
in the low-frequency regime is $p=5.8\alpha - 5.2$ (dashed)---this is not given correctly by either the synchrotron 
or Petrosian approximations. A spectral bump at $\chi = 0.6$ is characteristic of transrelativistic emission, and
is present for all $p>1.5$.
}
\end{figure}

\begin{figure}
\centering
\plotone{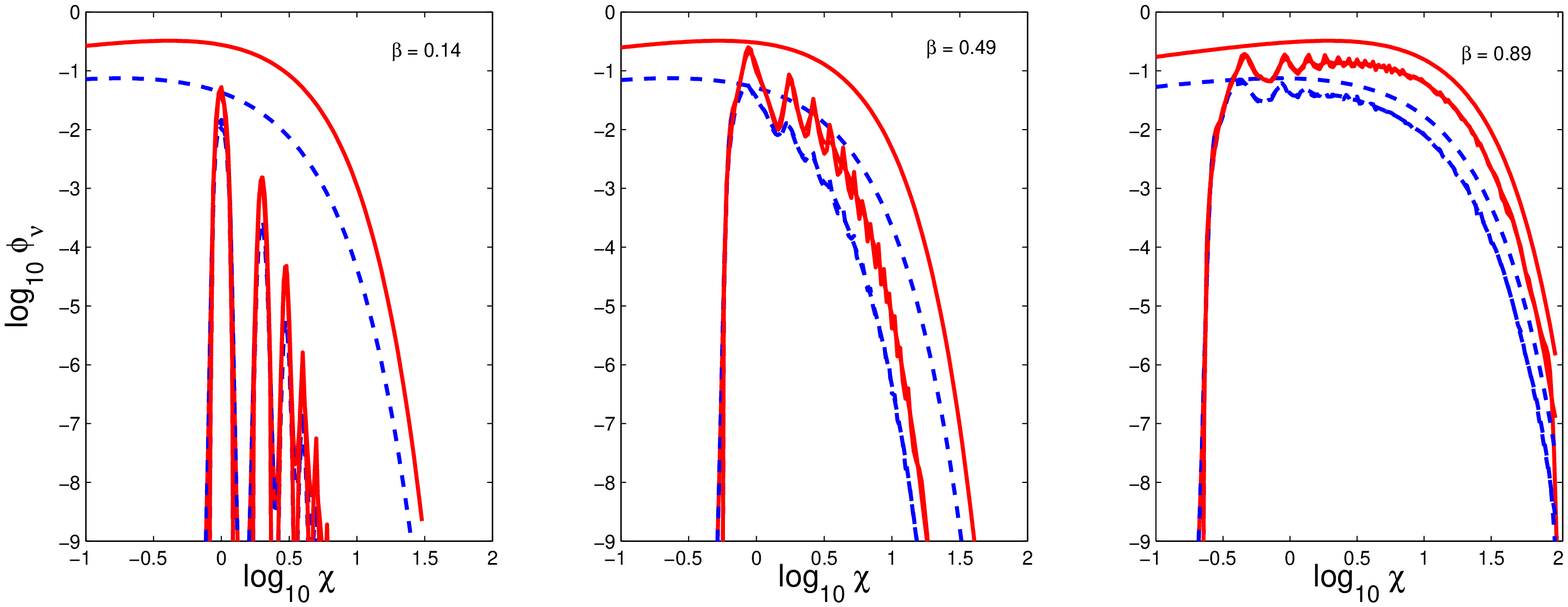}
\caption{
The emission is fit in the X (solid) and O (dashed) polarization directions separately, 
again between $0.1<\beta<0.95$, $ -1 < \log_{10}(\nu/\nu_b) < 2.0$; 
and again the mean error lies around $5\%$.
}
\end{figure}

\begin{figure}
\centering
\plotone{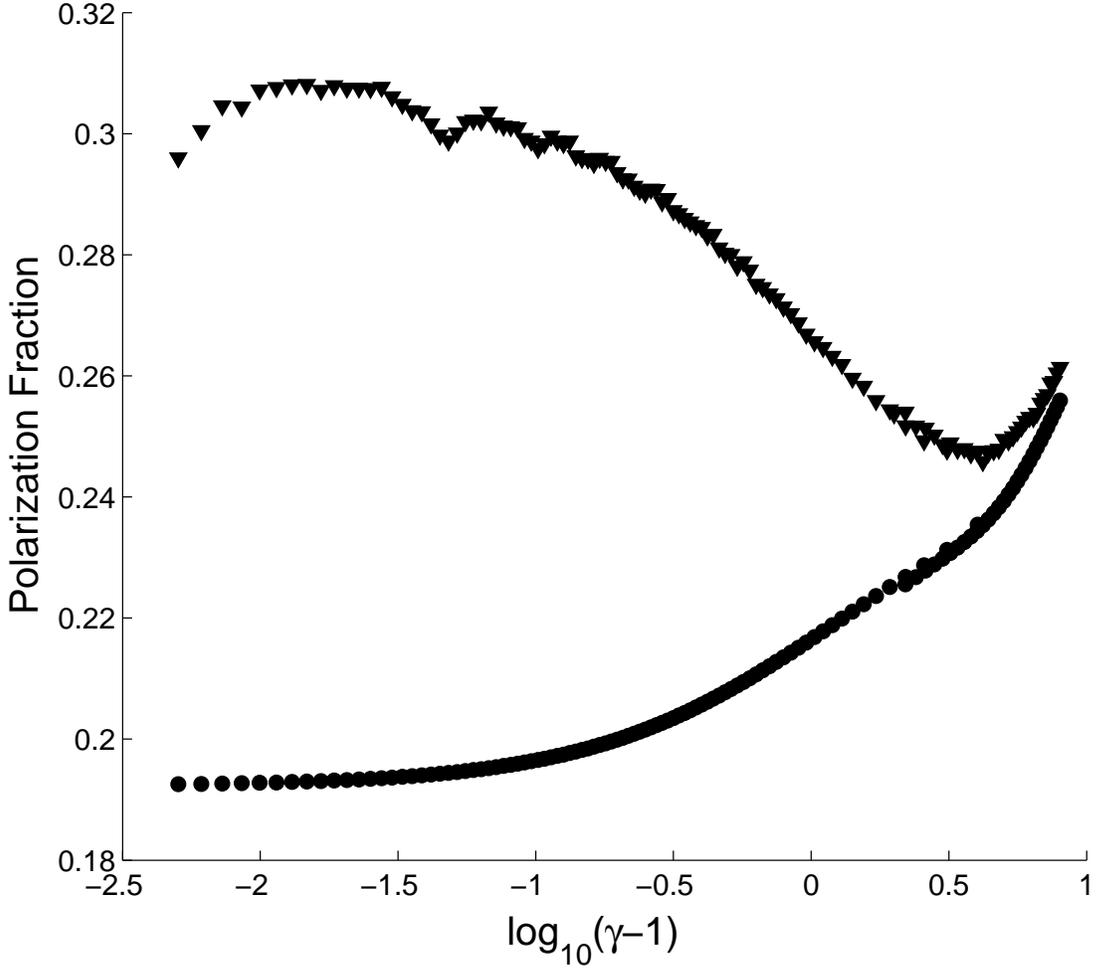}
\caption{
The frequency-integrated polarization fraction $\bar{\eta}_O/\bar{\eta}_X$
of a single particle (triangles) with $\beta \sim 0.5$  
is roughly $3/2$ that predicted by the relativistic theory (circles), between $-1.0<\log(\nu/\nu_b)<2.0$. 
In the low-energy limit it is dominated by the Trubnikov limit of $1/3$, while in the high-energy limit it
matches the synchrotron formula. Again, the synchrotron formula applies above $\gamma \sim 4$.
}
\end{figure}

\begin{figure}
\centering
\plotone{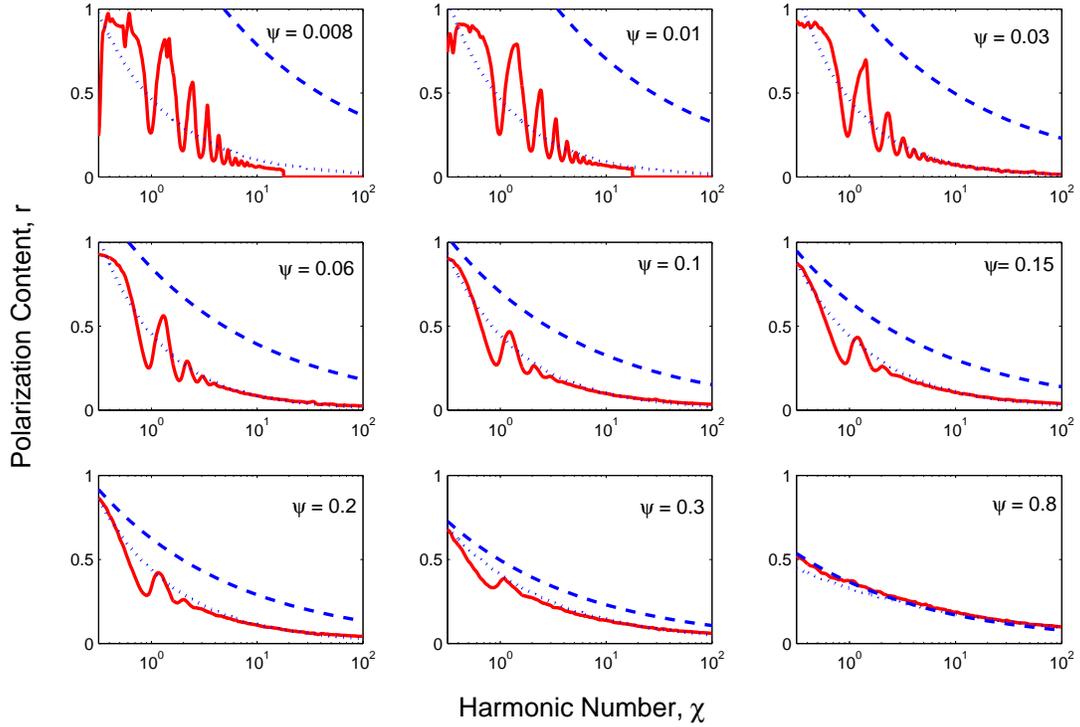}
\caption{
The polarization content, $r$, of a thermal distribution (solid), along with our fit (dotted). 
We have had difficulty matching the result of Dulk (1985), since it is unclear what assumptions he used
regarding the pitch angle distribution.  The approximation of Petrosian and McTiernan (1982) (dashed), 
valid at the peak of the emissivity and when $\psi \chi >> 1$, in fact agrees well for the more relaxed 
requirement $\psi > 0.5$. This is already quite near the synchrotron limit, however.  
}
\end{figure}

\begin{figure}
\centering
\plotone{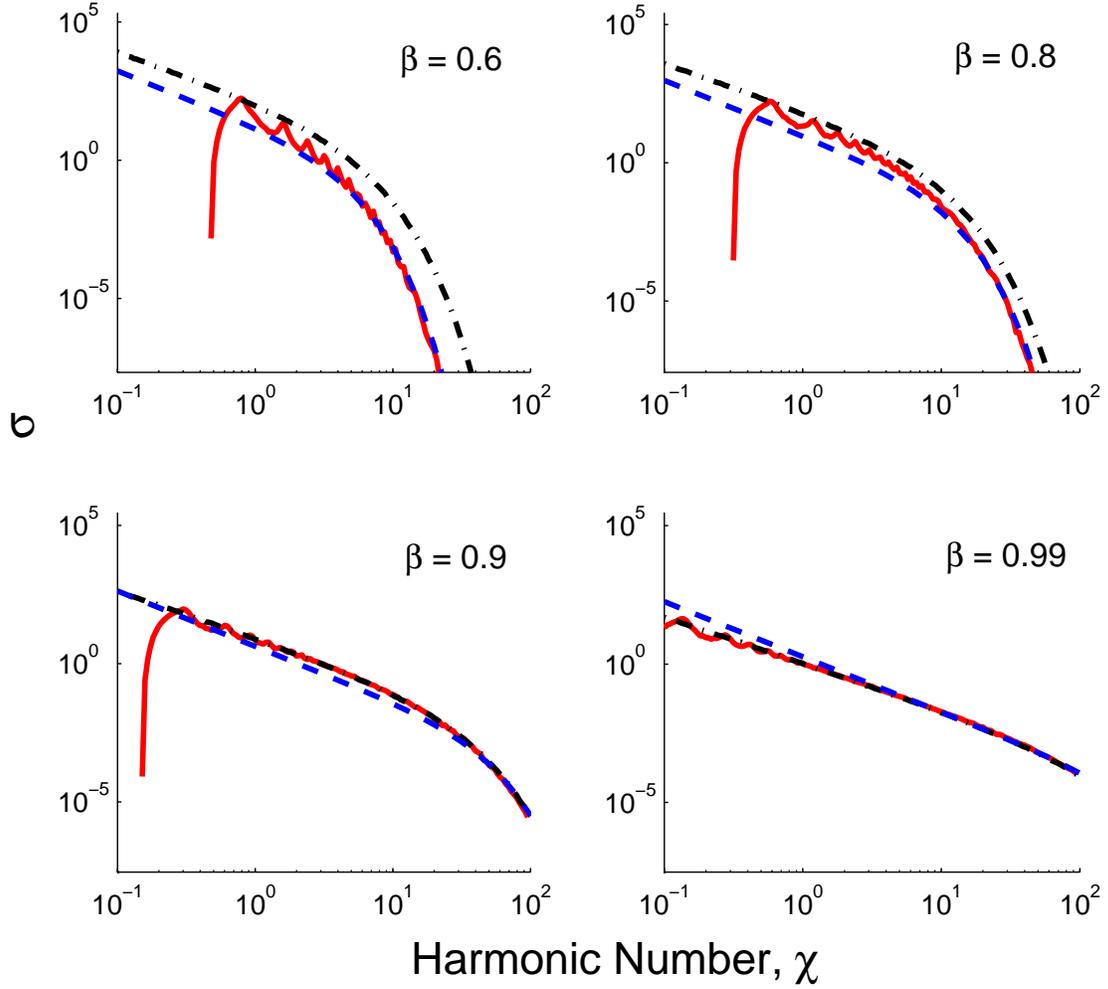}
\caption{
The cross-section from differentiation of Equation (2), together with the synchrotron limit and the angle-averaged
approximation of Petrosian (1981). As with the emissivity, the correct cross section in the low-frequency region is
roughly midway between the Petrosian and synchrotron approximations. The Petrosian approximation matches the
full integration for high frequencies, and in all cases the cross-section goes over to a power-law of index
$-5/3$ at high energies. As in Ghisellini \& Svensson (1991), we plot the dimensionless 
$\sigma_{Synch} = (\sigma_{Thomp} B_{cr} / \alpha_f B) \sigma$, where $\alpha_f$ is the fine-structure constant
and $B_{cr}\approx 4.4\times 10^{13}$ G is the criticial magnetic field above which quantum effects are important.
}
\end{figure}

\end{document}